\documentclass[a4paper,11pt]{article}
\usepackage{jheppub} 
\usepackage{lineno}
\usepackage{subfigure} 
\usepackage{float}
\usepackage{soul}
\usepackage{orcidlink}
\title{\boldmath Measurement of {\boldmath $B \to K{}^{*}(892)\gamma$} decays at Belle~II}

\collaboration{The Belle II Collaboration}
  \author{I.~Adachi\,\orcidlink{0000-0003-2287-0173},} 
  \author{L.~Aggarwal\,\orcidlink{0000-0002-0909-7537},} 
  \author{H.~Ahmed\,\orcidlink{0000-0003-3976-7498},} 
  \author{H.~Aihara\,\orcidlink{0000-0002-1907-5964},} 
  \author{N.~Akopov\,\orcidlink{0000-0002-4425-2096},} 
  \author{A.~Aloisio\,\orcidlink{0000-0002-3883-6693},} 
  \author{N.~Althubiti\,\orcidlink{0000-0003-1513-0409},} 
  \author{N.~Anh~Ky\,\orcidlink{0000-0003-0471-197X},} 
  \author{D.~M.~Asner\,\orcidlink{0000-0002-1586-5790},} 
  \author{H.~Atmacan\,\orcidlink{0000-0003-2435-501X},} 
% ...

 \collaboration{\includegraphics[height=17mm]{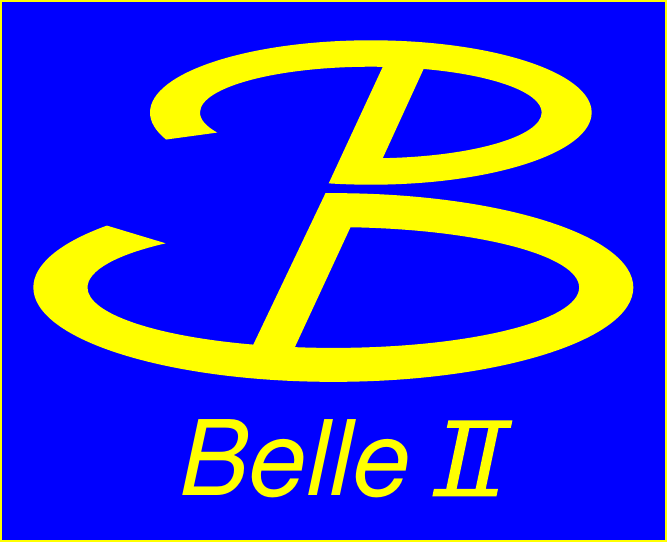}\\[6pt]
 The Belle~II collaboration}

\emailAdd{coll-publications@belle2.org}

\abstract{We present measurements of $B \to K{}^{*}(892)\gamma$ decays using $365\,{\rm fb}^{-1}$ of data collected from 2019 to 2022 by the Belle~II experiment at the SuperKEKB asymmetric-energy $e^+e^-$ collider. The data sample contains $(387 \pm 6) \times 10^6$ $\Upsilon(4S)$ events. We measure branching fractions ($\mathcal{B}$) and $C\!P$ asymmetries ($\mathcal{A}_{C\!P}$) for both $B^{0}\to K{}^{*0}\gamma$ and $B^{+}\to K{}^{*+}\gamma$ decays. The difference in $C\!P$ asymmetries ($\Delta \mathcal{A}_{C\!P}$) and the isospin asymmetry ($\Delta_{0+}$) between these neutral and charged channels are also measured. We obtain the following branching fractions and $C\!P$ asymmetries: $\mathcal{B} (B^{0} \to K{}^{*0}\gamma) = (4.14 \pm 0.10  \pm 0.11 ) \times 10^{-5}$, $\mathcal{B} (B^{+} \to K{}^{*+}\gamma) = (4.04 \pm 0.13  {}^{+0.13}_{-0.15} )\times 10^{-5}$, $\mathcal{A}_{C\!P} (B^{0} \to K{}^{*0}\gamma) = (-3.3 \pm 2.3  \pm 0.4 )\%$, and $\mathcal{A}_{C\!P} (B^{+} \to K{}^{*+}\gamma) = (-0.7 \pm 2.9  \pm 0.5 )\%$. The measured difference in $C\!P$ asymmetries is $\Delta \mathcal{A}_{C\!P} = (+2.6 \pm 3.8  \pm 0.6 )\%$, and the measured isospin asymmetry is $\Delta_{0+} = (+4.8 \pm 2.0  \pm 1.8 )\%$. The first uncertainties listed are statistical and the second are systematic. These results are consistent with world-average values and theory predictions.}

\begin{document}

\maketitle
\flushbottom

\section{Introduction}
\label{intro}

The radiative decays $B\to K{}^{*}(892)\gamma$ are suppressed in the Standard Model~(SM), where the dominant contribution comes from a one-loop $b\to s\gamma$ diagram~\cite{SM7, SM8}. Extensions to the SM predict new particles that can contribute to the internal loop, potentially altering branching fractions ($\mathcal{B}$) and other observables from their SM expectations.
Thus, these radiative decays are a promising probe for physics beyond the SM (BSM)~\cite{BSM1,BSM2}.
Observables for $B\to K{}^{*}(892)\gamma$ include the $C\!P$-violating asymmetry
\begin{equation}
    \mathcal{A}_{C\!P} \equiv \frac{\Gamma(\overline{B}\to\overline{K}{}^{*}\gamma) - \Gamma(B\to K{}^{*}\gamma)}{\Gamma(\overline{B}\to\overline{K}{}^{*}\gamma) + \Gamma(B\to K{}^{*}\gamma)}
\label{Eq:cpv-asy}
\end{equation}
and the isospin asymmetry
\begin{equation}
    \Delta_{0+} \equiv \frac{\Gamma(B^{0}\to K{}^{*0}\gamma) - \Gamma(B^{+}\to K{}^{*+}\gamma)}{\Gamma(B^{0}\to K{}^{*0}\gamma) + \Gamma(B^{+}\to K{}^{*+}\gamma)},
\label{Eq:iso-asy}
\end{equation}
where $\Gamma$ denotes the partial width and $K{}^{*}$ denotes the $K{}^{*}(892)$ meson. In Eq.~\ref{Eq:iso-asy} and throughout this paper, charge-conjugate modes are implicitly included unless otherwise noted.

The SM prediction for $\mathcal{B}(B\to K{}^{*}\gamma)$ has large uncertainties of about $30\%$, mostly because of form factors~\cite{SM1,SM2}.
In contrast, observables such as $\mathcal{A}_{C\!P}$ and $\Delta_{0+}$ have reduced uncertainties due to cancellation of form factor contributions and some experimental systematic uncertainties in the ratios of Eqs.~\ref{Eq:cpv-asy} and \ref{Eq:iso-asy}~\cite{SM3, SM4}.
In the SM, $\Delta_{0+}$ is estimated to have a small positive value ranging from around $3\%$~\cite{SM4} to $8\%$~\cite{SM5}. However, BSM scenarios such as the aligned two-Higgs-doublet model can shift $\Delta_{0+}$ to negative values as large as -7\%~\cite{BSM3}. The SM prediction for $\mathcal{A}_{C\!P}$ is less than $1\%$~\cite{BSM3, SM6}, while BSM physics can increase it.
The last measurement by the Belle experiment~\cite{Belle_paper} using $772\times 10^{6}$ $B\overline{B}$ events reported $\Delta_{0+} = (6.2 \pm 2.0)\%$ with a significance of $3.1$ standard deviations.
The BABAR experiment~\cite{BaBar_paper} also measured a positive $\Delta_{0+}$ value and set a 90\% confidence interval of $1.7\% < \Delta_{0+} < 11.6\%$.
For $\mathcal{A}_{C\!P}$, Belle, BABAR, and LHCb~\cite{LHCb_paper} have obtained values consistent with zero.
We report herein results for $\mathcal{B}$, $\mathcal{A}_{C\!P}$, and $\Delta_{0+}$ measured in $B\to K{}^{*}\gamma$ decays using $e^{+}e^{-}$ collision data recorded by the Belle~II experiment.

Subsequent sections of this paper are arranged as follows:
Section~\ref{Dataset} provides a brief introduction to the Belle~II experiment and the dataset.
Section~\ref{Reconstruction} describes the selection of final-state particles and $B$-meson reconstruction.
Section~\ref{CS} summarises the background suppression strategy. Section~\ref{Fit} describes the fit model and the procedure to calculate various observables. Section~\ref{Results} presents the results of the measurements and the fit projections.
Section~\ref{Systematics} describes the evaluation of systematic uncertainties.
Finally, Section~\ref{Summary} provides a summary of the results.

\section{The Belle~II Detector and Dataset}
\label{Dataset}

The Belle~II detector~\cite{belle2tdr} operates at the SuperKEKB accelerator~\cite{superkekb}, which collides $4\,{\rm GeV}$ positrons with $7\,{\rm GeV}$ electrons at center-of-mass (c.m.) energies at or near the $\Upsilon(4S)$ resonance.
Belle~II is arranged in a cylindrical geometry and features a two-layer silicon-pixel detector~\cite{Belle-IIDEPFET:2021pib} surrounded by a four-layer double-sided silicon-strip detector (SVD)~\cite{Belle-IISVD:2022upf} and a 56-layer central drift chamber (CDC), all used to reconstruct trajectories of charged particles (tracks). For the data analysed here, only one sixth of the second pixel layer was installed. The symmetry axis of these subdetectors, defined as the $z$ axis, is almost coincident with the electron beam direction. The $x$ axis is horizontal, with the positive direction pointing outward from the centre of the SuperKEKB storage ring, and the $y$ axis is vertical, with the positive direction upward. The polar and azimuthal angles are defined with respect to the positive $z$ and $x$ axes, respectively.
Surrounding the CDC, which also provides specific ionization measurements, are a time-of-propagation counter (TOP)~\cite{KOTCHETKOV2019162342} in the barrel region and an aerogel-based ring-imaging Cherenkov counter (ARICH) in the forward region.
These subdetectors are used for charged-particle identification (PID).
Surrounding the TOP and ARICH is an electromagnetic calorimeter (ECL) composed of CsI(Tl) crystals that provides energy and timing measurements for electrons and photons. Outside of the ECL, a superconducting solenoid generates a $1.5\,{\rm T}$ magnetic field oriented parallel to the $z$ axis.
The flux return of the solenoid is instrumented with resistive-plate chambers and plastic scintillators to detect muons and $K^{0}_{L}$ mesons.

The data used in this analysis were collected by Belle~II from 2019 to 2022 at a c.m.\ energy corresponding to the $\Upsilon(4S)$ resonance and $60\,{\rm MeV}$ below the resonance. The integrated luminosities for these on- and off-resonance datasets are $(365.3\pm 1.7)\,{\rm fb}^{-1}$ and $(42.7 \pm 0.2)\,{\rm fb}^{-1}$, respectively~\cite{luminosity}. The off-resonance dataset is used to study continuum background, i.e., $e^{+}e^{-}\to q\overline{q}$ ($q$ = $u, d, s, c$) events.
The analysis is performed in a ``blind" manner, in which all selection criteria and the fitting procedure are finalized before examining candidates in the signal region. We use Monte Carlo (MC) simulated events to optimize selection criteria, calculate reconstruction efficiencies, and study backgrounds. Simulated samples of $\Upsilon(4S)\to B\overline{B}$ events in which one of the $B$ mesons decays to the $K{}^{*}\gamma$ final state are used to study properties of the signal. Similarly, simulated $\Upsilon(4S)\to B\overline{B}$ events in which both $B$ mesons decay in an inclusive manner are used to study backgrounds from $B$ decays. All $\Upsilon(4S)\to B\overline{B}$ decays are generated with \textsc{EvtGen}~\cite{{EVTGEN}}. Continuum events are generated with \textsc{KKMC}~\cite{KKMC} interfaced to \textsc{Pythia}~\cite{Pythia}. The Belle~II detector response is simulated with \textsc{Geant4}~\cite{GEANT4}. Both simulated and real data samples are processed within the Belle~II software framework~\cite{basf2, basf2-zenodo}.

\section{Event Selection and Reconstruction}
\label{Reconstruction}

We reconstruct $B\to K{}^{*}\gamma$ decays proceeding via $K{}^{*0}\to K^{+}\pi^{-}$, $K{}^{*0}\to K^{0}_{S}\pi^{0}$, $K{}^{*+}\to K^{+}\pi^{0}$, and $K{}^{*+}\to K^{0}_{S}\pi^{+}$.
The reconstruction of pion and kaon tracks follows the algorithm described in Ref.~\cite{tracking}.
To ensure a reliable momentum measurement, we require at least 20 track hits in the CDC.
Tracks originating from near the $e^{+}e^{-}$ interaction point (IP) are required to satisfy ${d_{r} < 0.5~\rm cm }$ and ${|d_{z}| < 2.0~\rm cm }$, where $d_{r}$ ($d_{z}$) is the transverse (longitudinal) impact parameter of the track with respect to the IP.
To identify pion and kaon candidates, a PID likelihood is calculated based upon Cherenkov light measurements in the TOP and ARICH, and specific ionization measurements in the CDC and SVD. A track is identified as a pion if the ratio $\mathcal{L}(\pi)/[\mathcal{L}(K) + \mathcal{L}(\pi)] > 0.6$, where $\mathcal{L}(K)$ and $\mathcal{L}(\pi)$ are the likelihoods that a track is a kaon or pion, respectively. A track is identified as a kaon if $\mathcal{L}(K)/[\mathcal{L}(K) + \mathcal{L}(\pi)] > 0.6$. These criteria yield a $K^{+}$ ($\pi^{+}$) identification efficiency of approximately 79\% (77\%), with a probability of misidentifying a $\pi^{+}$ ($K^{+}$) as a $K^{+}$ ($\pi^{+}$) of around 10\%.

Candidates for high-energy photons coming directly from the $B$ decay are reconstructed from energy deposits (clusters) in the barrel and forward regions of the ECL, with energies satisfying $1.4\,\text{GeV} < E^{\ast}_{\gamma} < 3.4\,\text{GeV}$.
The asterisk ($\ast$) denotes quantities calculated in the $e^{+}e^{-}$ c.m.\ frame.
These clusters must have no matched tracks in the CDC.
To ensure their shape is consistent with an electromagnetic shower, we require that the ratio E9/E21 exceeds 0.9, where E9 and E21 are the energies deposited in a $3 \times 3$ array of crystals and a $5 \times 5$ array excluding the corners, respectively, centred on the crystal with the highest energy.
The difference between the ECL cluster time and the event time must be less than 200~ns to suppress out-of-time photons arising from beam backgrounds.
To distinguish high-energy photon candidates from $K^{0}_{L}$ clusters, we use a stochastic gradient-boosted decision tree (BDT)~\cite{FastBDT}.
The BDT is trained with 11 Zernike moments~\cite{Zernike} derived from the ECL shower shape.
These criteria result in a photon selection efficiency of $88\%$ and a background rejection of $85\%$.

We reconstruct $K^{0}_{S}$ candidates from pairs of oppositely charged tracks assumed to be pions.
We do not apply $d_{r}$, $d_{z}$, or PID selection criteria to these tracks, rather we fit them to a common vertex and require that the goodness-of-fit be satisfactory. The invariant mass of the $K^{0}_{S}$ candidate must be within $10\,{\rm MeV}\!/c^2$ of the known $K^{0}_{S}$ mass~\cite{PDG}.
Additional criteria are applied to these candidates, including momentum-dependent requirements on the $K^{0}_{S}$ flight distance in the transverse plane, the angle between the reconstructed $K^{0}_{S}$ momentum and the vector pointing from the IP to the $K^{0}_{S}$ vertex, and the distance along the $z$ axis from the IP to the $K^{0}_{S}$ vertex.
The $K^{0}_{S}$ selection efficiency is approximately $94\%$.

We reconstruct $\pi^{0}$ candidates from pairs of photons.
Each photon is required to have an energy greater than 80, 30, or 60 MeV, depending on whether it is detected in the forward, barrel, or backward region, respectively, of the ECL.
We require $\pi^{0}$ candidates to have a diphoton invariant mass ($m_{\gamma\gamma}$) in the range $120\,{\rm MeV}\!/c^2 < m_{\gamma\gamma} < 145\,{\rm MeV}\!/c^2$.
Further requirements are imposed on variables related to the photon candidates, i.e., ECL shower shape, polar and azimuthal angles, and the opening angle between the photons.
A mass-constrained fit is performed for the $\pi^{0}$ candidates to improve their momentum resolution.
The overall $\pi^0$ selection efficiency is approximately 75\%.

A $K{}^{*}$ candidate is formed from a $K^{+}\pi^{-}$, $K^{0}_{S}\pi^{0}$, $K^{+}\pi^{0}$, or $K^{0}_{S}\pi^{+}$ combination.
We retain $K{}^{*}$ candidates having an invariant mass within $75\,{\rm MeV}\!/c^2$ of the known $K{}^{*0}$ or $K{}^{*+}$ mass~\cite{PDG}.
This range corresponds to approximately 1.5 times the $K{}^{*}$ decay width.

We combine a $K{}^{*}$ with a high-energy photon candidate to form a $B$ candidate.
A vertex fit~\cite{TreeFit} is applied to the entire $B$ decay chain, with the $B$ production vertex constrained to the IP.
The $\chi^2$ probability of this vertex fit is required to be greater than $0.1\%$.
To identify signal decays, we define two kinematic variables: $M_{\rm bc}\equiv \sqrt{(E^{\ast}_{\rm beam})^{2}/c^{4}-(\vec{p}^{\,\ast}_{B})^{2}/c^{2}}$ and $\Delta E\equiv E^{\ast}_{B}-E^{\ast}_{\rm beam}$, where $E^{\ast}_{\rm beam}$ and $E^{\ast}_{B}$ are the beam and $B$ energies, respectively, and $\vec{p}^{\,\ast}_B$ is the $B$ momentum, all calculated in the c.m.\ frame.
To improve the $M_{\rm bc}$ resolution and reduce the correlation between $M_{\rm bc}$ and $\Delta E$, the $B$ momentum used here is
\begin{equation}
 \vec{p}^{\,\ast}_{B} = \vec{p}^{\,\ast}_{K\pi} + \frac{\vec{p}^{\,\ast}_{\gamma}}{|\vec{p}^{\,\ast}_{\gamma}|} \times (E^{\ast}_{\rm beam}-E^{*}_{K\pi}),  
\end{equation}
i.e., the magnitude of the photon momentum in $M_{\rm bc}$ is taken to be the difference $(E^{\ast}_{\rm beam}-E^{*}_{K\pi})$.
The improvement in the $M_{\rm bc}$ resolution is around $16\%$ ($6\%$) for channels without (with) a $\pi^{0}$ in the final state.
We retain events satisfying $5.23\,{\rm GeV}\!/c^{2}< M_{\rm bc} < 5.29\,{\rm GeV}\!/c^{2}$ and $|\Delta E|< 0.30\,{\rm GeV}$ for further analysis.
To calculate signal yields, we define a narrower signal region $M_{\rm bc}>5.27\,{\rm GeV}\!/c^{2}$ and $-0.15\,{\rm GeV} < \Delta E < 0.07\,{\rm GeV}$; these ranges correspond to $\pm 3\sigma$ in resolution.
The asymmetric region for $\Delta E$ accounts for energy leakage from the ECL crystals.
All selection criteria are optimized with a figure-of-merit (FOM) calculated as $N_S/\sqrt{N_S+N_B}$, where $N_S$ and $N_B$ are the number of simulated signal and background events within the signal region.

\section{Background Suppression}
\label{CS}
Large backgrounds arise from photons produced in decays of high-momentum $\pi^{0}$ and $\eta$ mesons.
To reduce these backgrounds, we implement $\pi^{0}$ and $\eta$ vetoes as follows.
We pair the primary photon candidate from the signal $B$ decay with other photons in the event and reject the signal candidate if the photon pair is consistent with arising from a $\pi^{0}$ or $\eta$ decay as determined by a BDT classifier.
Separate classifiers are used for $\pi^{0}$ and $\eta$ mesons.
The photon with which the signal photon is paired must have a minimum energy: for a $\pi^{0}$ ($\eta$) candidate, the energy must be greater than 20~MeV (30~MeV).
In the forward region where beam backgrounds are higher, the energy must be greater than 25~MeV (35~MeV).
The classifiers use as inputs the diphoton invariant mass and various quantities for the low-energy photon: its energy, polar angle, ECL shower shape, distance between the ECL cluster and tracks extrapolated from the CDC, the ratio $E9/E21$, and the Zernike-based BDT output.
The $\pi^{0}$-veto BDT also uses the cosine of the angle in the putative $\pi^{0}$ rest frame between the signal photon and the boost direction of the lab frame.
We require that the BDT outputs exceed certain thresholds, which are determined by maximizing an FOM.
These requirements reject approximately 64\% of background photons while preserving 89\% of signal decays.

The remaining background is dominated by $e^{+}e^{-}\to q\overline{q}$ continuum events, which are characterized by a back-to-back jet topology.
In contrast, because $B\overline{B}$ events are produced almost at rest in the $e^{+}e^{-}$ c.m.\ frame, they have a nearly isotropic distribution of final-state particles.
To suppress continuum background, we employ another BDT classifier referred to as CSBDT.
This classifier uses the following inputs: modified Fox--Wolfram moments~\cite{fox_wolfram}, variables characterizing the momentum flow about the signal candidate's thrust axis, the $\chi^2$ of the fit for the $B$ decay vertex, the distance along the $z$ axis between the $B$ signal vertex and the tag-side $B$ vertex, and the output of the $B$ flavour tagger~\cite{flavour}.
More details on these variables are given in Refs.~\cite{Bevan:2014iga,Belle_II_Physics}.
A criterion is applied to the CSBDT output that maximizes an FOM, separately for each $K{}^{*}$ decay channel.
This selection rejects 70--80\% of continuum background while preserving 84--95\% of signal decays, depending on the $K{}^{*}$ channel.

After applying all selection criteria, 0.4--5.6\% of events have multiple signal candidates, depending on the $K{}^{*}$ decay channel.
For such events, we retain the candidate with the highest CSBDT output value.
According to MC simulation, this choice selects the correctly reconstructed signal decay in events with multiple candidates 62--75\% of the time.

Decays of $B$ mesons into multibody final states accompanied by a high-energy photon are a potential background.
These decays typically exhibit a peaking structure in $M_{\rm bc}$, as they originate from a $B$ meson.
However, they tend to have a $\Delta E$ distribution shifted towards a negative value, away from the expected signal peak.
Simulation studies show that these backgrounds arise from a number of sources, e.g., $B\to K{}^{*}\pi^{0}$ and $B\to K{}^{*}\eta$ decays in which a photon from the subsequent $\pi^{0}$ or $\eta$ decay evades the veto; and decays to higher kaonic resonances, such as $B\to K_{1}(1270)\gamma$, $B\to K_{1}(1400)\gamma$, and $B\to K_{2}^{*}(1430)\gamma$. 
We combine contributions from all non-signal $B$ decays, including multibody final states with high-energy photons, into a single $B\overline{B}$ background component.

\section{Fit strategy}
\label{Fit}

We determine the signal yield from a two-dimensional extended maximum-likelihood fit to the unbinned $M_{\rm bc}$ and $\Delta E$ distributions.
The fit incorporates three components: signal, continuum, and $B\overline{B}$ background.
The probability density function (PDF) for each component is determined by fitting simulated events.
The likelihood function is written as
\begin{equation}
\mathcal{L}(\vec{\alpha})=\frac{e^{-\sum_{j} n_{j}}}{N!}\prod^{N}_{i=1}\sum^{}_{j} n_{j}\times\mathcal{P}_{j}(M_{\rm bc}^{i},\Delta E^{i};{\vec{\alpha}}),    
\label{likel}
\end{equation}
where $n_j$ and $\mathcal{P}_j$ are the number of events and PDF for component $j$, and $N$ is the total number of events.
The argument $M_{\rm bc}^{i}$ ($\Delta E^{i}$) denotes the $M_{\rm bc}$ ($\Delta E$) value for candidate $i$, and $\vec{\alpha}$ are the shape parameters for the PDF. The likelihood function is maximized with respect to $n_j$ and several PDF shape parameters, as described below.

Given the low correlation between $M_{\rm bc}$ and $\Delta E$ (less than 5\%), we use the product of their one-dimensional PDFs in the fit.
For signal events, the $M_{\rm bc}$ distribution is described by a Crystal Ball (CB) function~\cite{CB}.
The CB mean and width parameters are floated in the fit, while parameters describing the power-law tail are fixed to values obtained from simulation.
The $\Delta E$ distribution is modelled with a double-sided CB function.
The mean parameter of this function is also floated in the fit.
We modify the two width parameters using a common multiplicative factor, i.e., $\sigma_{L} \to f_{\Delta E}\times\sigma_{L}$ and $\sigma_{R} \to f_{\Delta E}\times\sigma_{R}$, where $\sigma_{L}$ and $\sigma_{R}$ are fixed to values obtained from simulation.
The $f_{\Delta E}$ parameter is floated in the fit.
All other shape parameters are fixed to values obtained from simulation.
By floating the mean and introducing a common multiplicative factor for the width, we account for potential data--simulation differences in the $\Delta E$ distribution of signal events. 
Up to 9\% of the signal component comes from misreconstructed $B\to K^{*}\gamma$ decays, referred to as ``self-crossfeed".
The self-crossfeed component is modelled with a two-dimensional nonparametric PDF~\cite{KDE}.
The total signal PDF is the sum of the PDFs for correctly reconstructed $B\to K{}^{*}\gamma$ decays and self-crossfeed events, with the latter fraction fixed to the value from simulation.

For continuum background, the $M_{\rm bc}$ and $\Delta E$ distributions are modelled with an ARGUS function~\cite{ARGUS} and a first-order polynomial, respectively. We use the product of these one-dimensional PDFs in the fit. The $B\overline{B}$ background is modelled with a two-dimensional nonparametric PDF~\cite{KDE}.
Except for the $B^{0}\to K{}^{*0}[K^{0}_{S}\pi^{0}]\gamma$ channel, the flavour of the $B$ meson is identified based on the charge of the hadrons from the $K{}^{*}$ decays. For example, a $K^{+}$ indicates a $B^{0}\to K{}^{*0}[K^{+}\pi^{-}]\gamma$ decay. For these decay modes, separate data samples are prepared for $B$ and $\overline{B}$ candidates. The branching fraction and $\mathcal{A}_{C\!P}$ for these decays are calculated from the result of a simultaneous fit to the $B$ and $\overline{B}$ samples as
\begin{equation}		
\mathcal{B} = \frac{N_{S}/\epsilon_{S} + N_{\overline{S}}/\epsilon_{\overline{S}} }{2 \times N_{\Upsilon(4S)} \times f_{+-} (f_{00})},
\label{BF1}
\end{equation}
\begin{equation}
\mathcal{A}_{C\!P} = \frac{N_{S}/\epsilon_{S} - N_{\overline{S}}/\epsilon_{\overline{S}}}{N_{S}/\epsilon_{S} + N_{\overline{S}}/\epsilon_{\overline{S}}},
\label{ACP}
\end{equation}
where $N_{S}$ ($N_{\overline{S}}$) is the fitted signal yield for a given $B\to K{}^{*}\gamma$ ($\overline{B}\to \overline{K}{}^{*}\gamma$) decay channel, $\epsilon_{S}$ ($\epsilon_{\overline{S}}$) is the corresponding selection efficiency, $N_{\Upsilon(4S)}$ is the number of $\Upsilon(4S)$ events, and $f_{+-} = 0.5113^{+0.0073}_{-0.0108}$ ($f_{00} = 0.4861^{+0.0074}_{-0.0080}$) is the branching fraction for $\Upsilon(4\text{S})\to B^{+}B^{-} (B^{0}\overline{B}{}^{0})$~\cite{HFLAV_2024}. The shape parameters for signal are common for the $B$ and $\overline{B}$ samples, whereas the shape parameters for continuum and $B\overline{B}$ backgrounds are floated separately for the two samples.
The $\mathcal{A}_{C\!P}$ calculated from Eq.~\ref{ACP} includes a small contribution due to interactions of charged hadrons with detector material.
This instrumental asymmetry, discussed in Section~\ref{Systematics}, is subtracted from the obtained $\mathcal{A}_{C\!P}$ value.

For the $B^{0}\to K{}^{*0}[K^{0}_{S}\pi^{0}]\gamma$ channel, the branching fraction is calculated as:
\begin{equation}
\mathcal{B} = \frac{N_{B^{0}\to K{}^{*0}[K^{0}_{S}\pi^{0}]\gamma}/\epsilon}{2\times N_{\Upsilon(4S)}\times f_{00}},
\label{BF2}
\end{equation}	
where $\epsilon$ is the selection efficiency for $B^{0}\to K{}^{*0}[K^{0}_{S}\pi^{0}]\gamma$.
We combine the branching fractions and $C\!P$ asymmetries for all charged and all neutral channels by calculating their weighted average, taking into account correlations among the systematic uncertainties~\cite{BLUE_1}.

The isospin asymmetry $\Delta_{0+}$, and the difference in $C\!P$ asymmetries $\Delta \mathcal{A}_{C\!P}$ between the neutral and charged $B\to K{}^{*}\gamma$ channels, are calculated from the relations
\begin{equation}
 \Delta_{0+} = \frac{(\tau_{+}/\tau_{0})\times\mathcal{B}(B^{0}\to K{}^{*0}\gamma) - \mathcal{B}(B^{+}\to K{}^{*+}\gamma)}{(\tau_{+}/\tau_{0})\times\mathcal{B}(B^{0}\to K{}^{*0}\gamma) + \mathcal{B}(B^{+}\to K{}^{*+}\gamma)}, 
\label{Del0+}
\end{equation}	
\begin{equation}
\Delta \mathcal{A}_{C\!P} = \mathcal{A}_{C\!P}(B^{+}\to K{}^{*+}\gamma) - \mathcal{A}_{C\!P}(B^{0}\to K{}^{*0}\gamma),
\label{DelACP}
\end{equation}	
where $\tau_{+}$ and $\tau_{0}$ are the known lifetimes~\cite{PDG} of $B^{+}$ and $B^{0}$, respectively.

\section{Results}
\label{Results}

\begin{figure}[H] 
\begin{center}
	\subfigure	
	{
	\includegraphics[width=0.3\paperwidth]{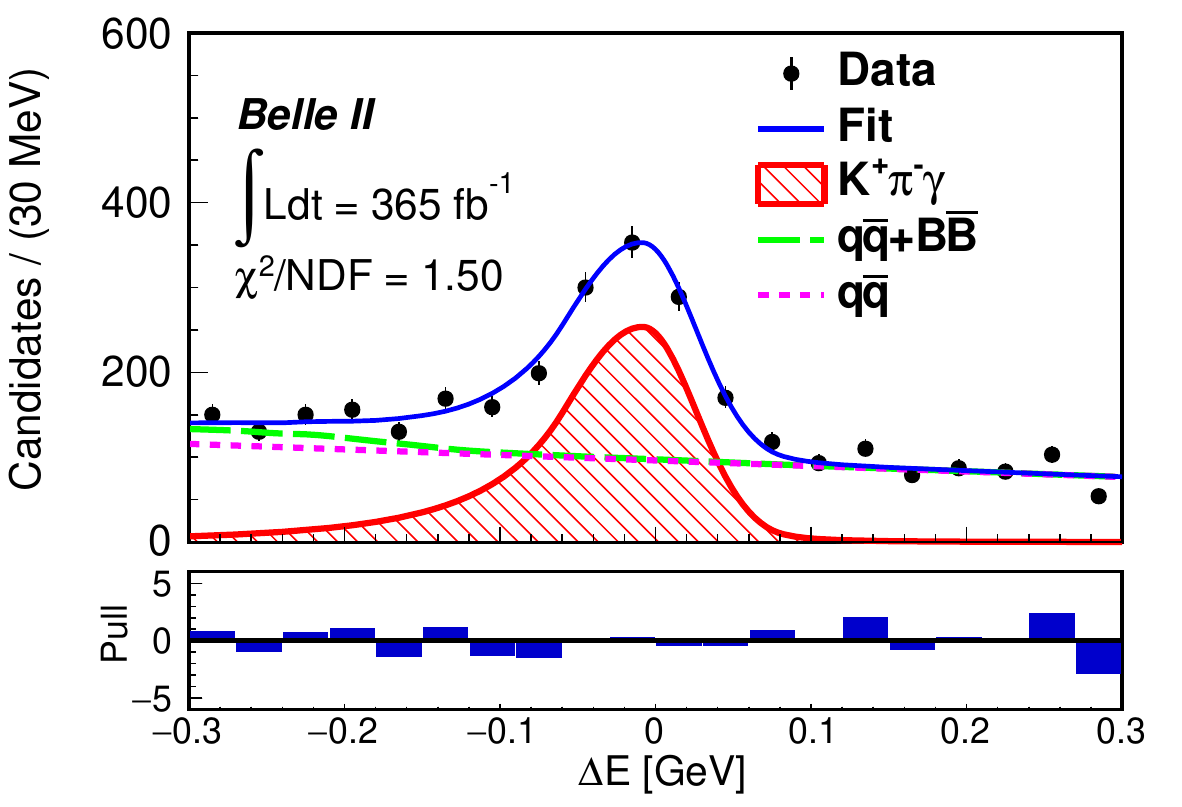}
	}
        \subfigure	
	{
	\includegraphics[width=0.3\paperwidth]{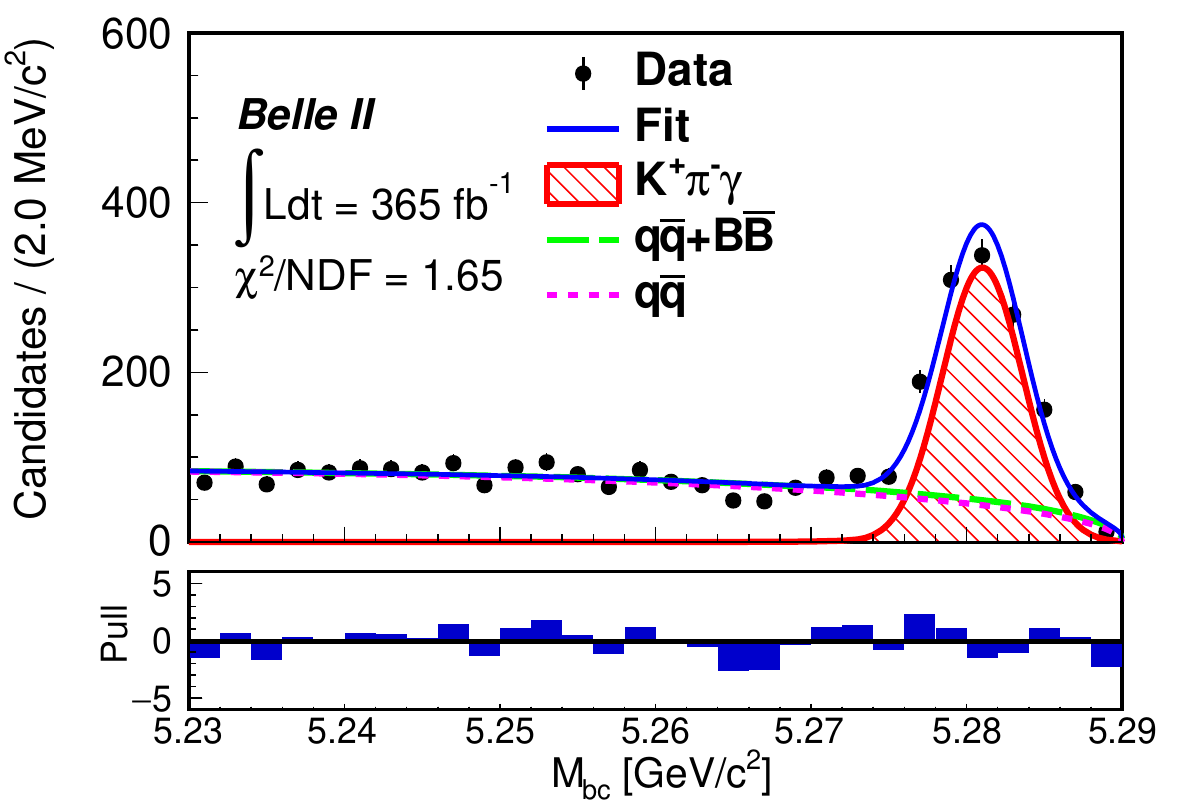}
	}
	\subfigure	
	{
	\includegraphics[width=0.3\paperwidth]{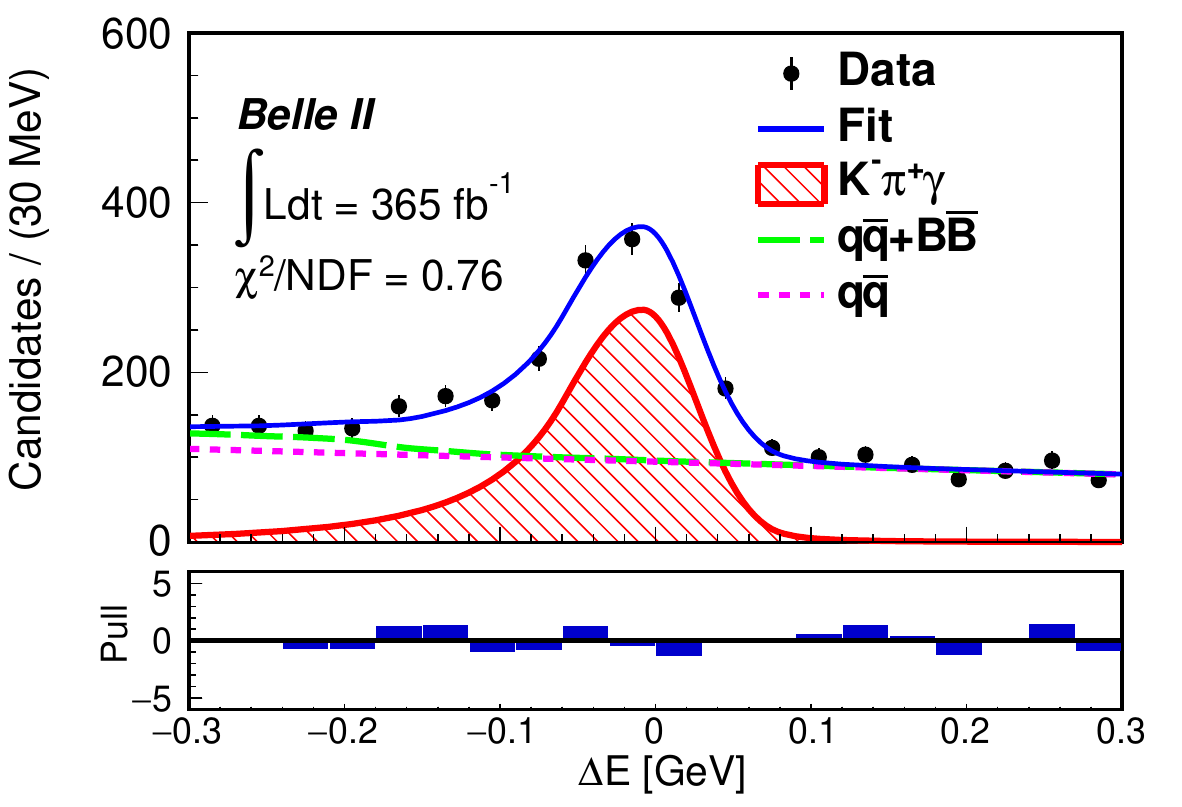}
	}
	\subfigure
        {
	\includegraphics[width=0.3\paperwidth]{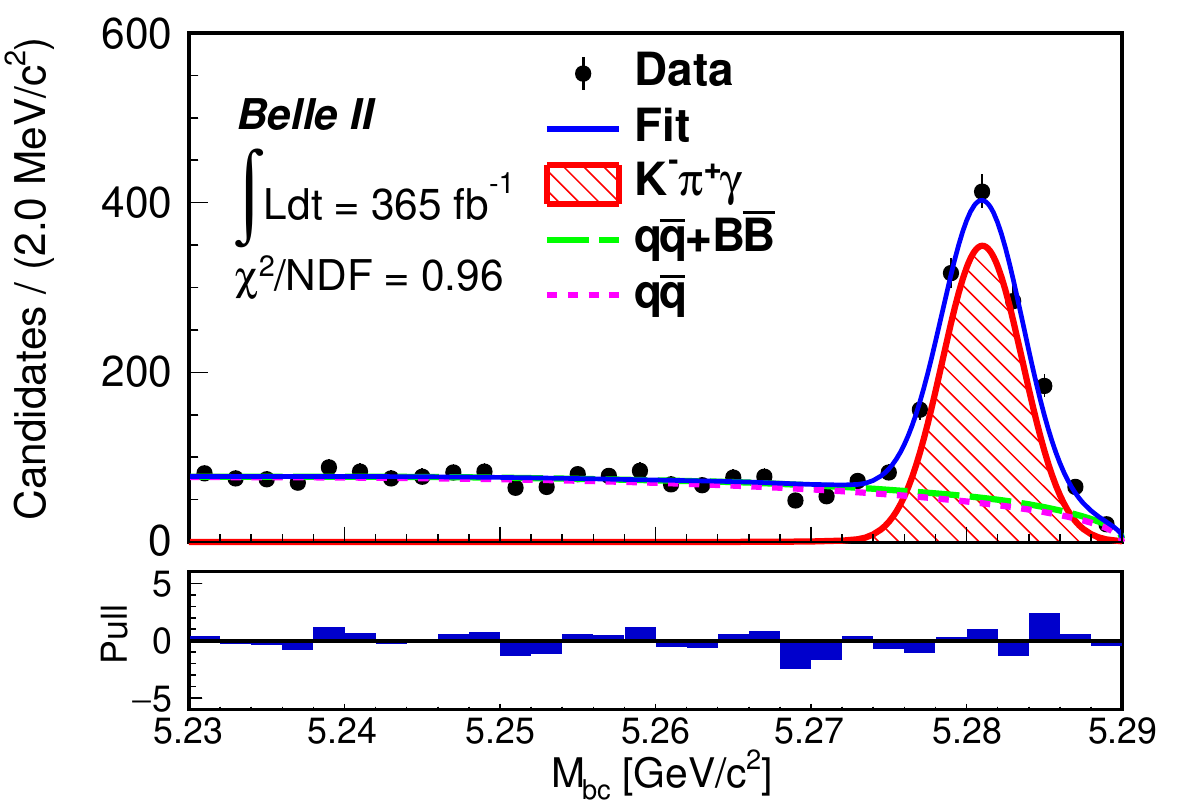}
	}
        \subfigure
	{
	\includegraphics[width=0.3\paperwidth]{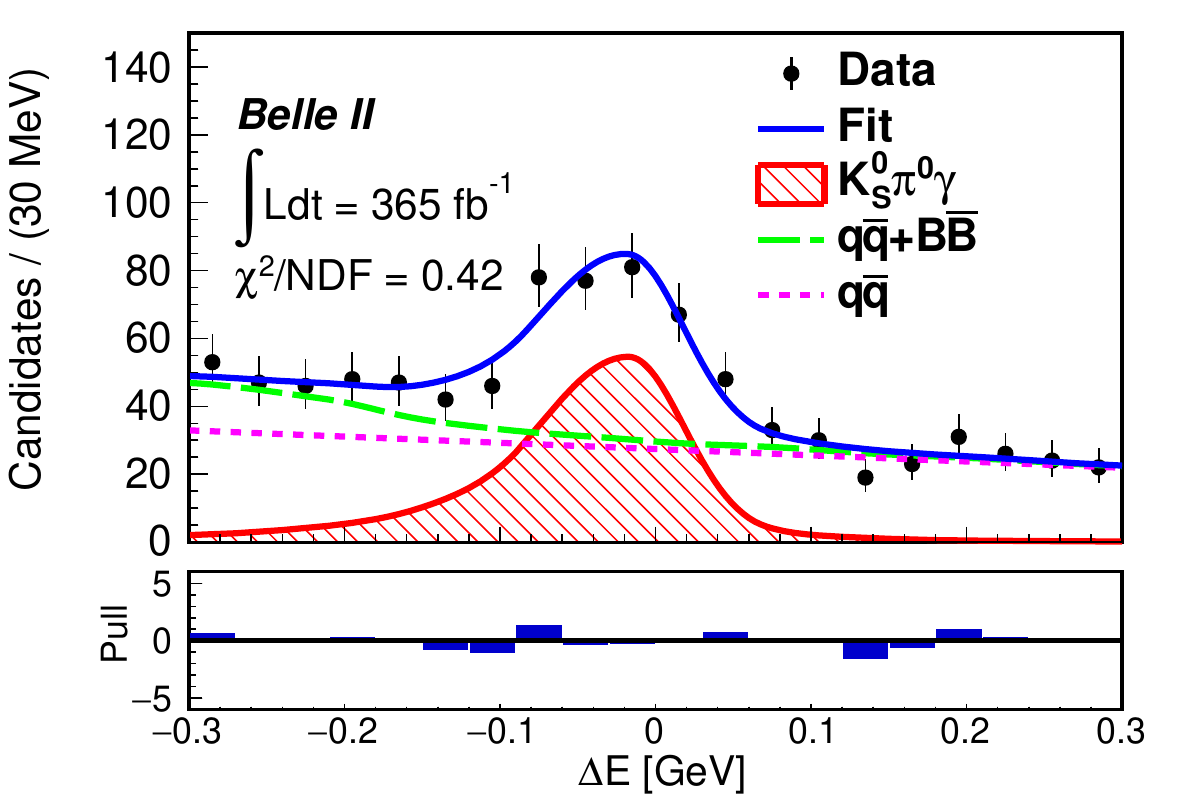}
	}
	\subfigure
        {
	\includegraphics[width=0.3\paperwidth]{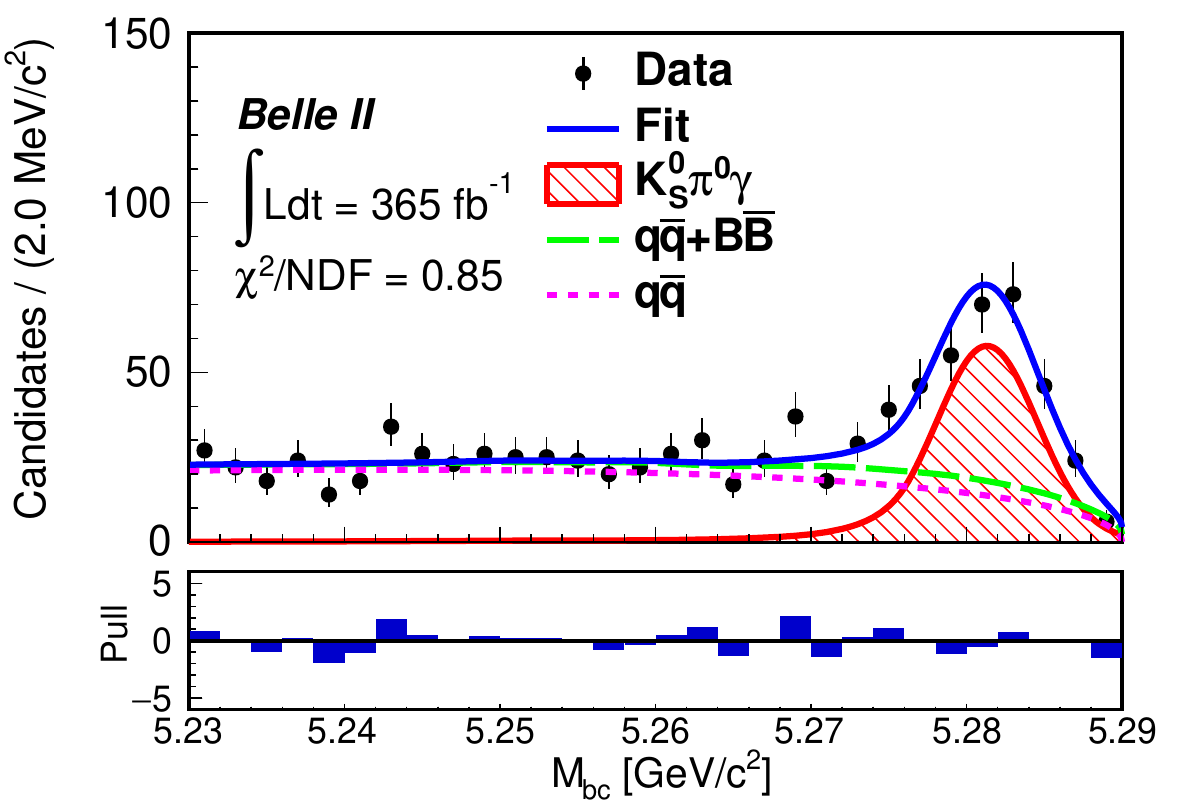}
	}
	\caption{The $\Delta E$ (left) and $M_{\rm bc}$ (right) distributions for $B^{0}\to K{}^{*0}\gamma$ channels with fit results superimposed. The top two rows and bottom row are for the $K^{\pm} \pi^{\mp} \gamma$ and $K^{0}_{S} \pi^{0} \gamma$ final state, respectively. The black dots with error bars are the data, the blue curves show the total fit, the filled red curves show the signal, the dashed green curves show the continuum background, and the dotted magenta curves show the $B\overline{B}$ background. Lower panels show the pull, i.e., the differences between the data and fit results divided by the statistical uncertainty of the data.}
	\label{Fig:fit_result1}
 \end{center}
 \end{figure}

\begin{figure}[H] 
\begin{center}
        \subfigure	
	{
	\includegraphics[width=0.3\paperwidth]{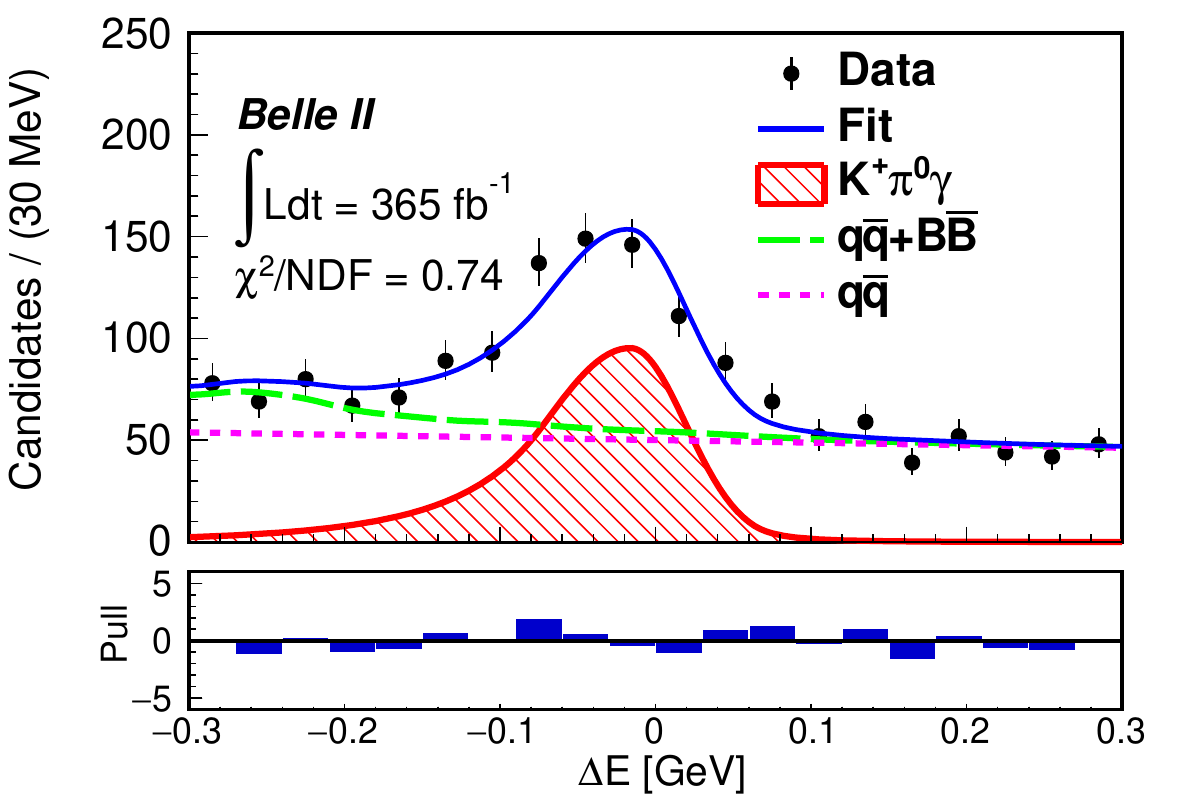}
	}
	\subfigure
    {
	\includegraphics[width=0.3\paperwidth]{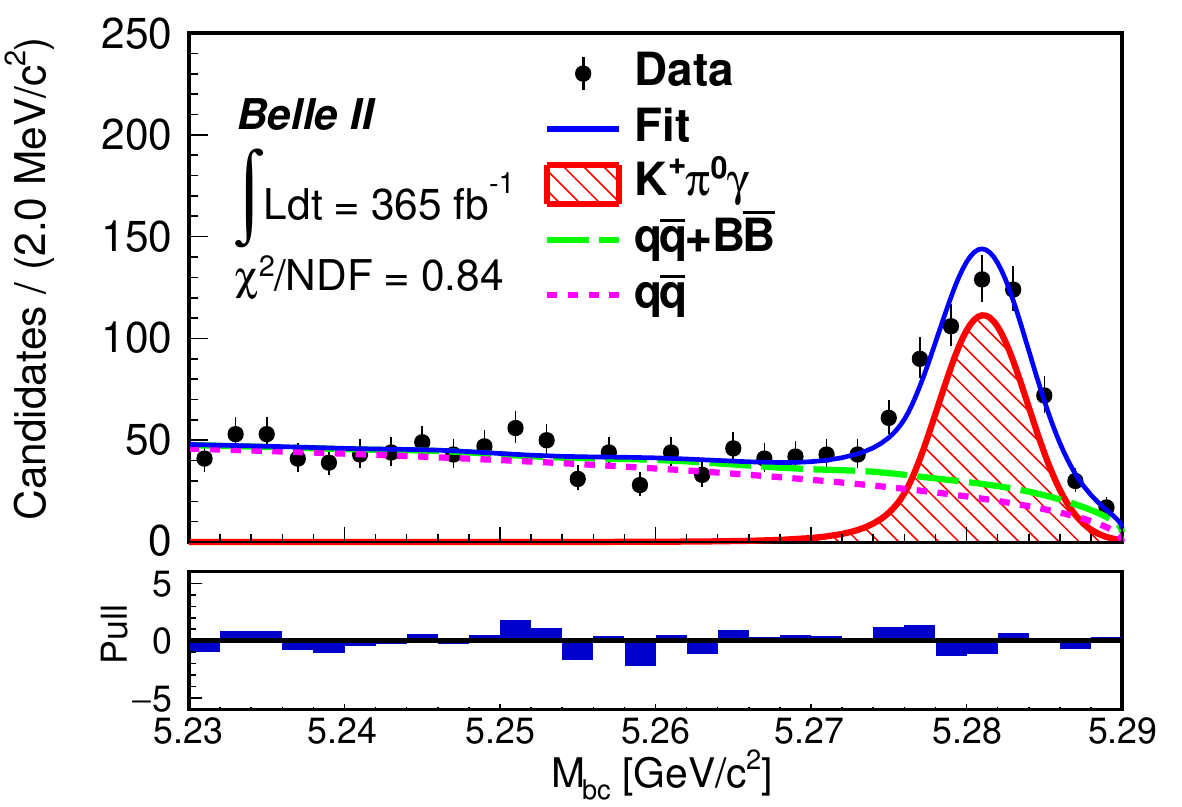}
	}
	\subfigure	
	{
	\includegraphics[width=0.3\paperwidth]{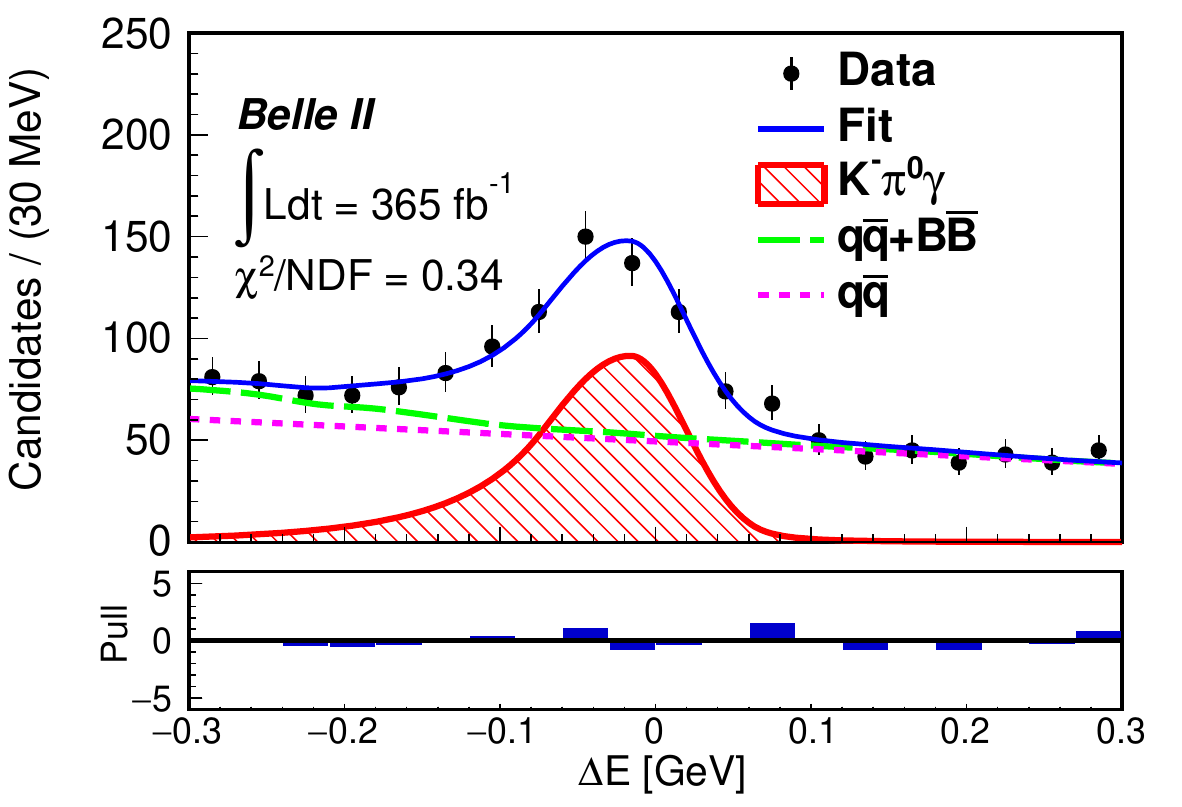}
	}
	\subfigure	
	{
	\includegraphics[width=0.3\paperwidth]{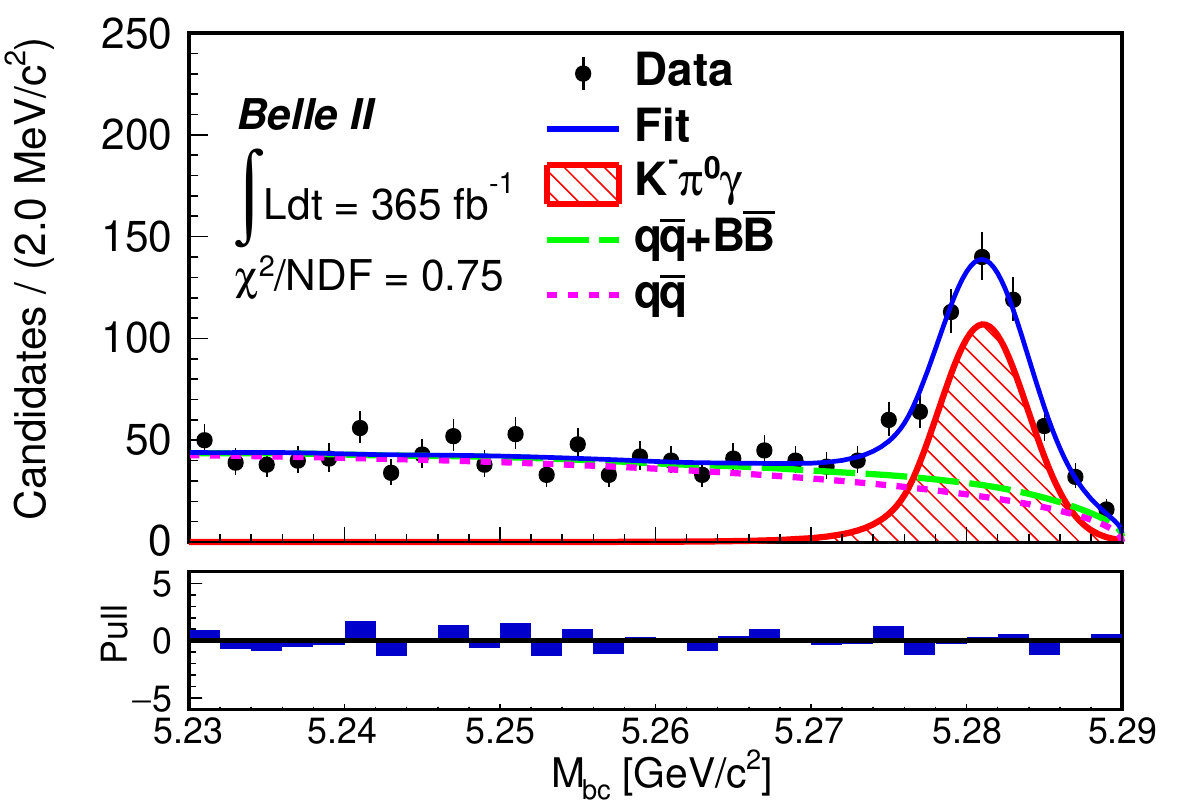}
	}
        \subfigure	
	{
	\includegraphics[width=0.3\paperwidth]{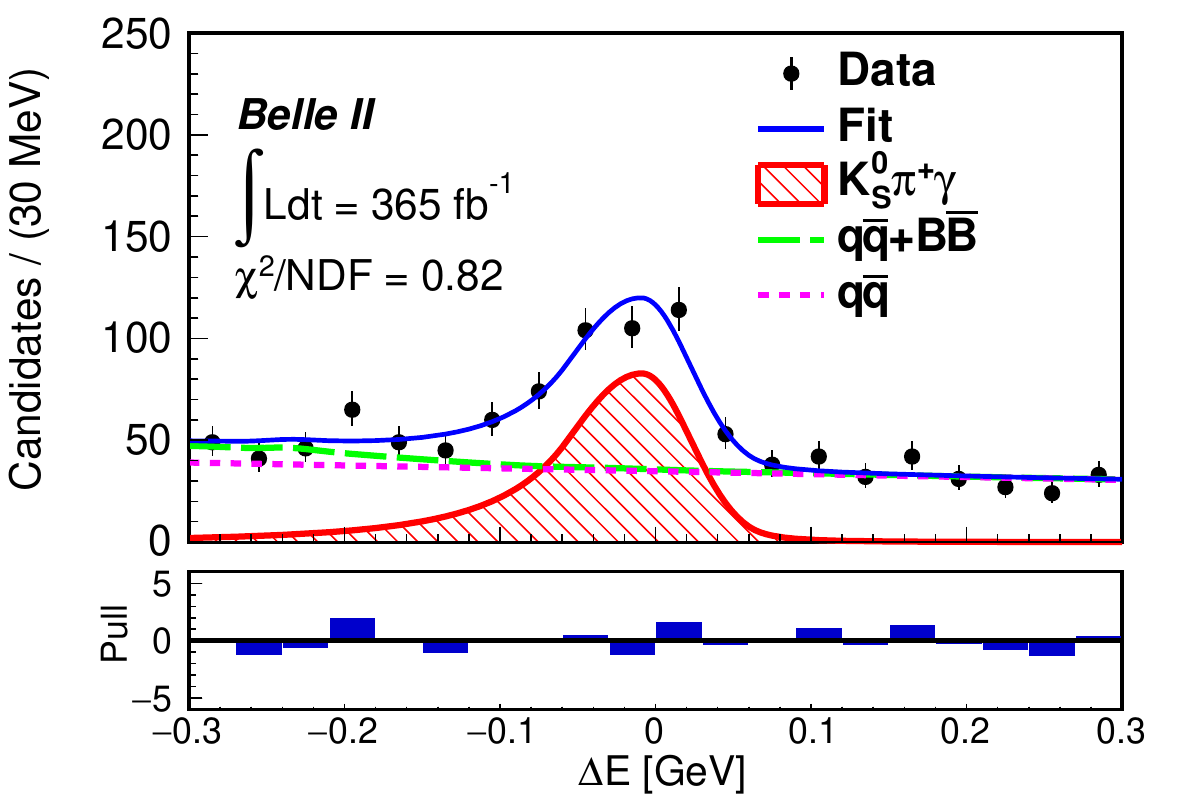}
	}
	\subfigure
        {
	\includegraphics[width=0.3\paperwidth]{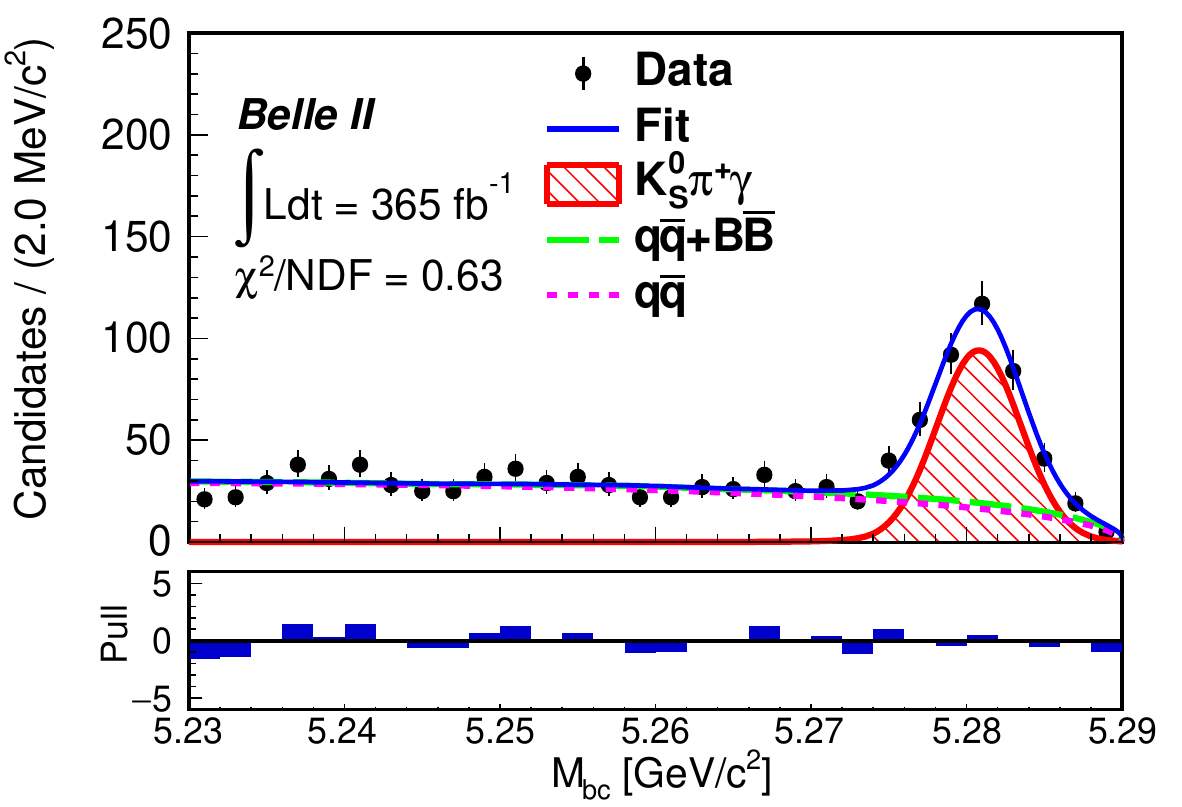}
	}
	\subfigure	
	{
	\includegraphics[width=0.3\paperwidth]{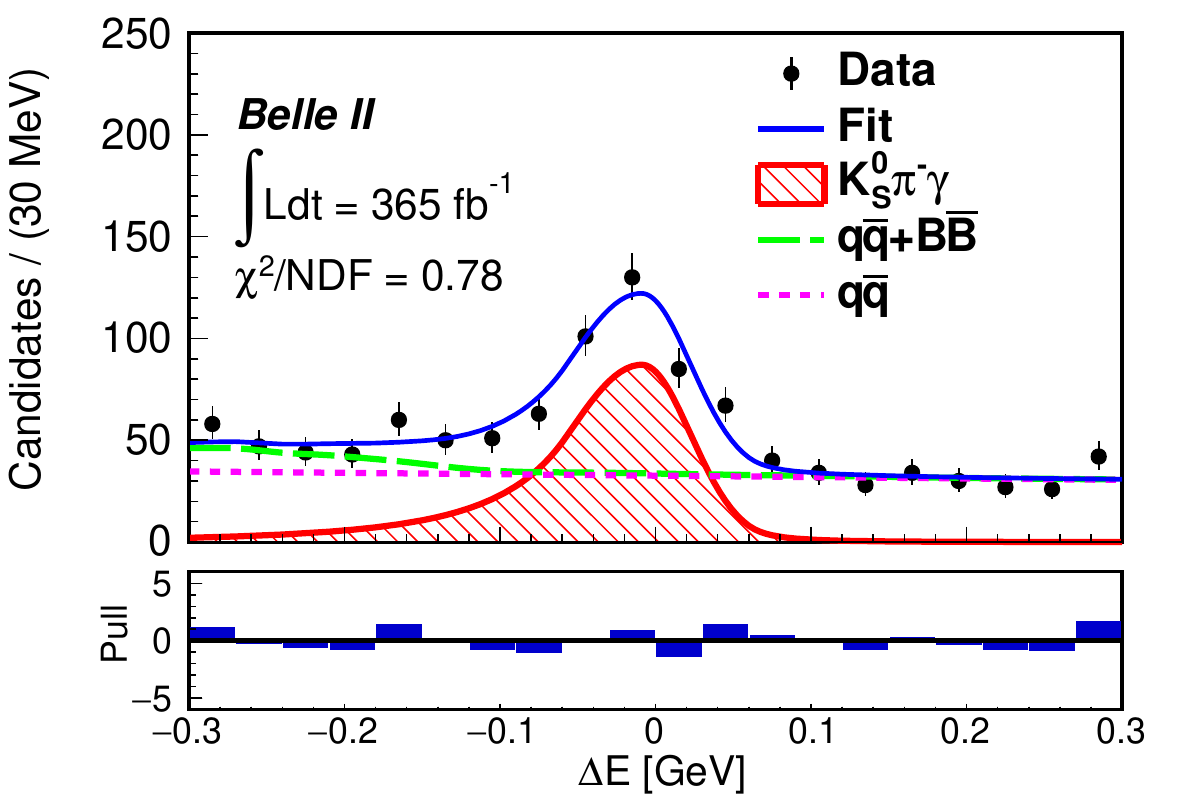}
	}
	\subfigure	
	{
	\includegraphics[width=0.3\paperwidth]{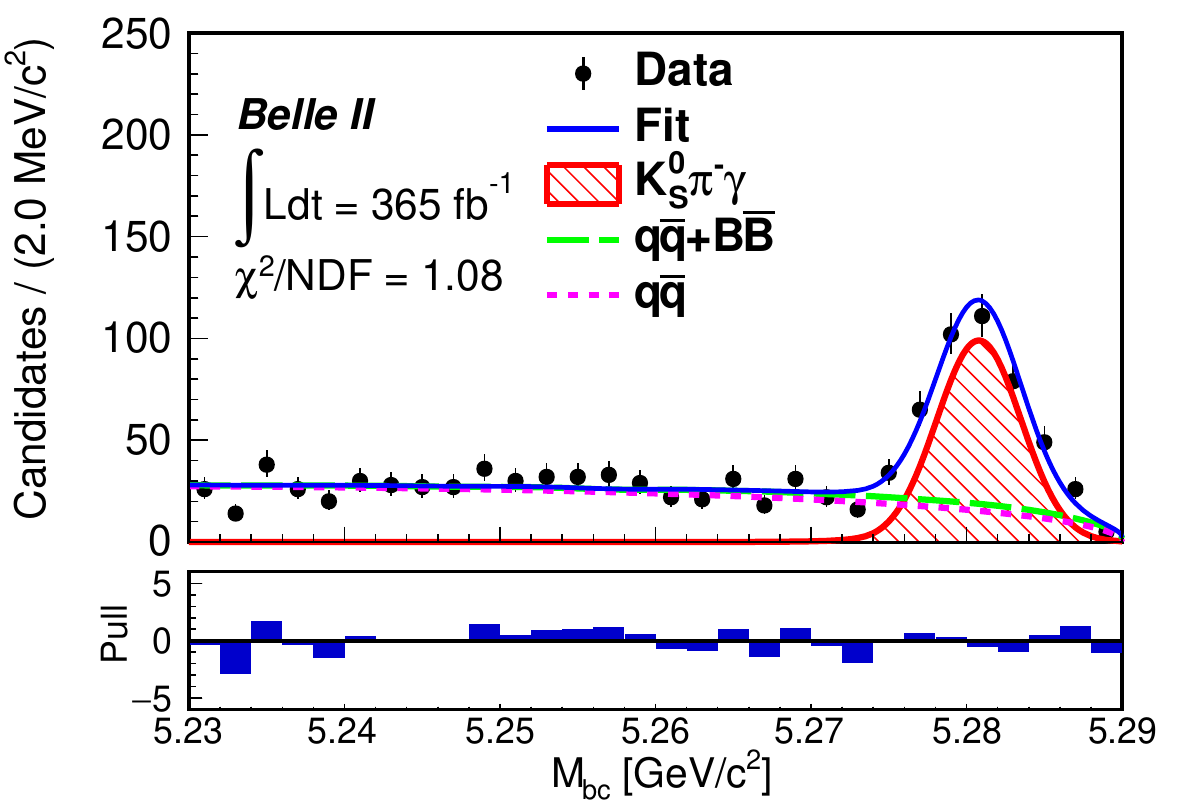}
	}
	\caption{The $\Delta E$ (left) and $M_{\rm bc}$ (right) distributions of charged ${B\to K{}^{*}\gamma}$ channels with the fit results superimposed. The top two rows and bottom row are for the $K^{\pm} \pi^{0} \gamma$ and $K^{0}_{S} \pi^{\pm} \gamma$ final state, respectively. The black dots with error bars are the data, the blue curves show the total fit, the filled red curves show the signal, the dashed green curves show the continuum background, and the dotted magenta curves show the $B\overline{B}$ background. Lower panels show the pull, i.e., the differences between the data and fit results divided by the statistical uncertainty of the data.}
	\label{Fig:fit_result2}
 \end{center}
 \end{figure}

The selection efficiencies and the yield of $B\to K{}^{*}\gamma$ signal, continuum ($N_{q\overline{q}}$) and $B\overline{B}$ backgrounds ($N_{b\overline{b}}$) obtained from the fits are listed in Table~\ref{Tab:fit_result}.
The results for $\mathcal{B}$, $\mathcal{A}_{C\!P}$, $\Delta_{0+}$, and $\Delta\mathcal{A}_{C\!P}$ calculated with Eqs.~\ref{BF1}--\ref{DelACP} are listed in Table~\ref{Tab:obs_result}.
The systematic uncertainties quoted are discussed in the next section.
The $M_{\rm bc}$ and $\Delta E$ distributions for the final event samples are shown in Figs.~\ref{Fig:fit_result1}--\ref{Fig:fit_result2}, along with projections of the fit result.
All results for $\mathcal{A}_{C\!P}$ and $\Delta\mathcal{A}_{C\!P}$ are consistent with zero, while $\Delta_{0+}$ differs from zero by {2.0} standard deviations.
As a check, we use the ${}_{s}\mathcal{P}lot$ method~\cite{Pivk:2004ty} to obtain the background-subtracted helicity distribution of the $K\pi$ system in the data, and we find it to be consistent with the distribution obtained from simulated $B\to K{}^{*}\gamma$ decays.

\begin{table}[htb!]
    \centering
        \caption{{Selection efficiencies and fitted yields of signal, $q\overline{q}$, and $B\overline{B}$ background events for each decay channel.
        Except for $B^{0}\to K^{*0}[K^{0}_{S}\pi^{0}]\gamma$ channel, the efficiencies listed for other channels are the average selection efficiencies $(\epsilon_{S} + \epsilon_{\overline{S}})/2$.
        The uncertainties listed for efficiencies and yields are systematic and statistical, respectively.
        The former are discussed in Section~\ref{Systematics}.}}
        \label{Tab:fit_result}
        \begin{tabular}{ c c c c c c}
			\hline
			Channel & {$\epsilon$ (\%)} & {$N_{S}$} & {$N_{\overline{S}}$} & $N_{q\overline{q}}$ & $N_{b\overline{b}}$\\
			\hline
			$B^{0}\to K{}^{*0}[K^{+}\pi^{-}]\gamma$ & {$14.26\pm 0.21$} & {$1068 \pm 36$} & {$1153 \pm 38$} & $3818 \pm 118$ & $188 \pm 76$\\
			$B^{0}\to K{}^{*0}[K^{0}_{S}\pi^{0}]\gamma$ & {$1.66\pm 0.09$} & \multicolumn{2}{c}{$254 \pm 20$} & $548 \pm 46$ & $86 \pm 40$\\
			$B^{+}\to K{}^{*+}[K^{+}\pi^{0}]\gamma$ & {$5.62\pm 0.26$} & {$453\pm27$}	& {$435\pm26$} & $1991 \pm 107$ & $221 \pm 92$\\
			$B^{+}\to K{}^{*+}[K^{0}_{S}\pi^{+}]\gamma$& {$4.22\pm 0.10$} & {$332\pm21$} &	{$349\pm21$} & $1348 \pm 75$ & $105 \pm 53$\\
			\hline		
	\end{tabular}
\end{table}

\begin{table}[htb!]
    \centering
        \caption{Measured branching fractions, $\mathcal{A}_{C\!P}$, $\Delta_{0+}$, and $\Delta\mathcal{A}_{C\!P}$ for $B\to K{}^{*}\gamma$.
        The first uncertainties listed are statistical and the second are systematic; the latter are discussed in Section~\ref{Systematics}.
        The third uncertainty on $\Delta_{0+}$ is due to the $f_{+-}/f_{00}$ ratio.}
        \label{Tab:obs_result}
        \begin{tabular}{ c c c }
			\hline
			Channel & $\mathcal{B}~(10^{-5})$ & $\mathcal{A}_{C\!P}~(\%)$\\
			\hline
$B^{0}\to K^{*0}[K^{+}\pi^{-}]\gamma$	&	$4.14 \pm 0.10 \pm 0.11$	&	$-3.3 \pm 2.3 \pm 0.4$	\\
$B^{0}\to K^{*0}[K^{0}_{S}\pi^{0}]\gamma$	&	$4.07 \pm 0.33 \pm 0.23$ 	&	$-$	\\
\hline					
$B^{0}\to K^{*0}\gamma$	&	$4.14 \pm 0.10 \pm 0.11$  	&	$-3.3 \pm 2.3 \pm 0.4$	\\
\hline					
$B^{+}\to K^{*+}[K^{+}\pi^{0}]\gamma$	&	$3.99 \pm 0.17 {}^{+0.20}_{-0.21}$  	&	$+1.7 \pm 4.0 \pm 0.8$	\\
$B^{+}\to K^{*+}[K^{0}_{S}\pi^{+}]\gamma$	&	$4.08 \pm 0.18 {}^{+0.13}_{-0.14}$ 	&	$-3.5 \pm 4.3 \pm 0.6$	\\
\hline					
$B^{+}\to K^{*+}\gamma$	&	$4.04 \pm 0.13 {}^{+0.13}_{-0.15}$ 	&	$-0.7 \pm 2.9 \pm 0.5$	\\
\hline					
    & $\Delta_{0+}$ (\%) & $\Delta \mathcal{A}_{C\!P}$ (\%)\\			
\hline					
$B\to K^{*} \gamma$	&	$+4.8 \pm 2.0 \pm 1.0 \pm 1.5$ 	&	$+2.6 \pm 3.8 \pm 0.6$	\\
   
			\hline		
	\end{tabular}
\end{table}

\section{Systematic Uncertainties}
\label{Systematics}

The various sources of systematic uncertainty are listed in Tables~\ref{table:Systematic} and~\ref{tab:Acp_sys} and discussed below. The uncertainties that are identical for $B$ and $\overline{B}$ decays, such as the number of $B\overline{B}$ events, photon selection, or $\pi^{0}/\eta$ veto, do not contribute to the $\mathcal{A}_{C\!P}$ measurement.

A systematic uncertainty of $1.5\%$ is assigned due to the number of $\Upsilon(4S)$ events, and an additional systematic contribution of $1.6\%$ ($2.1\%$) is assigned due to $f_{00}$ ($f_{+-}$). The systematic uncertainty due to track reconstruction is obtained from studies of $e^{+}e^{-} \to \tau^{+}\tau^{-}$ events. We assign a systematic uncertainty of 0.2\% per track, which results in an uncertainty of 0.5\% for the $K{}^{*0}[K^{+}\pi^{-}]$ and $K{}^{*0}[K^{0}_{S}\pi^{0}]$ channels, 0.2\% for $K{}^{*+}[K^{+}\pi^{0}]$, and 0.7\% for $K{}^{*+}[K^{0}_{S}\pi^{+}]$. The difference in efficiency between data and simulation due to PID selection criteria is studied with $D^{*+} \to D^{0}[K^{-}\pi^{+}]\pi^{+}$ decays. These corrections are calculated in bins of momentum and cosine of the polar angle of the track. We assign a systematic uncertainty of $0.2\%$ ($0.4\%$) for pion (kaon) selection. The uncertainty in the selection for the high-energy photon is 0.9\%, calculated from a control sample of $e^{+}e^{-}\to \mu^{+}\mu^{-}\gamma$ events. 

\begin{table}[htb!]
\centering
 \caption{Systematic uncertainties (\%) for the branching fractions.}
  	\label{table:Systematic}
  \begin{tabular}{ l c c c c }
		\hline
        Source &$K{}^{*0}[K^{+}\pi^{-}]\gamma$ & $K{}^{*0}[K^{0}_{S}\pi^{0}]\gamma$ & $K{}^{*+}[K^{+}\pi^{0}]\gamma$ & $K{}^{*+}[K^{0}_{S}\pi^{+}]\gamma$ \\
        \hline   
Number of $B\overline{B}$ events	&	1.6	&	1.6	&	1.6	&	1.6	 \\
$f_{+-}/f_{00}$	&	${}^{+1.4}_{-1.6}$	&	${}^{+1.4}_{-1.6}$	&${}^{+1.4}_{-2.1}$ &${}^{+1.4}_{-2.1}$	\\
Tracking efficiency	&	0.5	&	0.5	&	0.2	&	0.7	\\
$\pi^{+}$ selection	&	0.2	&	$-$	&	$-$	&	0.2	\\
$K^{+}$ selection	&	0.4	&	$-$	&	0.4	&	$-$	\\
$\gamma$ selection 	&	0.9	&	0.9	&	0.9	&	0.9	\\
$K^{0}_{S}$ reconstruction	&	$-$	&	1.4	&	$-$	&	1.4	\\
$\pi^{0}$ reconstruction	&	$-$	&	3.9	&	3.9	&	$-$	\\
$\chi^{2}$ requirement	&	0.2	&	1.0	&	0.2	&	1.0	\\
$\pi^{0}$ veto	&	0.7	&	0.7	&	0.7	&	0.7	 \\
$\eta$ veto	&	0.2	&	0.2	&	0.2	&	0.2	\\
CSBDT requirement	&	0.3	&	0.4	&	0.4	&	0.3	\\
Best candidate selection	&	0.1	&	1.0	&	0.5	&	0.2	\\
Fit bias	&	0.1	&	0.3	&	0.2	&	0.2	\\
Signal PDF model	&	0.2 &	0.4	&	0.3	&	0.2	\\
$B\overline{B}$ PDF model	&	0.2	&	0.4	&	0.4	&	0.3		\\
Simulation sample size 	&	0.2	&	0.8	&	0.4	&	0.5		\\
Self-crossfeed fraction &	0.2	&	2.2	&	1.8	&	0.5	\\
\hline
Total	&	${}^{+2.6}_{-2.7}$	&	${}^{+5.6}_{-5.7}$ &	${}^{+5.0}_{-5.3}$	&	${}^{+3.2}_{-3.5}$	\\
        \hline
        		\end{tabular} 
\end{table}

The systematic uncertainty of $K^{0}_{S}$ reconstruction is estimated with a $D^{+}\to K^{0}_{S}\pi^{+}$ control sample.
The $K^{0}_{S}$ and charged pion are reconstructed with identical selection criteria as the signal channel.
To reconstruct $D^{+}$ candidates, $K^{0}_{S}$ and $\pi^{+}$ candidates are combined and required to have an invariant mass satisfying $|M_{K^{0}_{S}\pi^{+}} - m_{D^{+}}| < 20\,{\rm MeV}\!/c^2$, where $m_{D^{+}}$ is the known $D^{+}$ mass~\cite{PDG}.
The momentum of the $K^{0}_{S}$ candidate is restricted to the range $0.5\,{\rm GeV}\!/c < p < 3.0\,{\rm GeV}\!/c$, to match the signal channel.
By comparing the yields of $K^{0}_{S}$ as a function of the flight distance in data and simulation, we estimate the systematic uncertainty due to the $K^{0}_{S}$ selection to be 1.4\%.

The systematic uncertainty for $\pi^{0}$ selection is obtained from studies of $D^{0} \to K^{-}\pi^{+}\pi^{0}$ and $D^{0} \to K^{-} \pi^{+}$ decays.
A systematic uncertainty of 3.9\% is assigned for the $\pi^{0}$ selection. The dominant contributions are uncertainties in the branching fractions of $D^{0}$ control modes. 

\begin{table}[htb!]
\centering
    \caption{Systematic uncertainties (\%) for $\mathcal{A}_{C\!P}$. }
	\label{tab:Acp_sys}
		\begin{tabular}{ l c c c }
			\hline
        Source &$K{}^{*0}[K^{+}\pi^{-}]\gamma$ &$K{}^{*+}[K^{+}\pi^{0}]\gamma$ &$K{}^{*+}[K^{0}_{S}\pi^{+}]\gamma$ \\
        \hline
Fit bias	&	0.1	&	0.1 &	0.1	\\
Signal PDF model	&	0.1	&	0.1	&	0.1	\\
$B\overline{B}$ PDF model	&	0.1	&	0.4	&	0.2	\\
Best candidate selection	&	0.1	&	0.5	&	0.2	\\
$K^{+}$ asymmetry	&	$-$	&	0.5	&	$-$	\\
$\pi^{+}$ asymmetry	&	$-$	&	$-$	&	0.5	\\
$K^{+}\pi^{-}$ asymmetry	&	0.3	&	$-$	&	$-$	\\
\hline
Total	&	0.4	&	0.8	&	0.6	\\
			\hline
		\end{tabular}
\end{table}

The difference in efficiency between data and simulation for vertex-quality selection criteria is studied with $\overline{B}{}^{0}\to D^{+}[\to K^{0}_{S}\pi^{+}]\pi^{-}$ and $B^{-}\to D^{0}[\to K^{-}\pi^{+}]\pi^{-}$ control samples. These $b\to c$ channels have relatively large branching fractions and low background contamination. The results from $\overline{B}{}^{0} \to D^{+}[K^{0}_{S}\pi^{+}]\pi^{-}$ and $B^{-} \to D^{0}[K^{-}\pi^{+}]\pi^{-}$ are used for decays with and without, respectively, a $K^{0}_{S}$ in the final state. The selection criteria for these $B\to D\pi$ decays mirror those for the signal $B\to K{}^{*}\gamma$ channels. We select $D^{+}$ and $D^{0}$ candidates within an invariant-mass window $|M_{K\pi} - m_{D}| < 10\,{\rm MeV}\!/c^2$, where $m_{D}$ is the known $D$ mass~\cite{PDG}.
The $D$ meson is then paired with a prompt pion to form a $B$ candidate. As for the signal channel, the range of $M_{\rm bc}$ is restricted to $5.23\,{\rm GeV}\!/c^2 < M_{\rm bc} < 5.29\,{\rm GeV}\!/c^2$. However, since there is no photon in the final state, $\Delta E$ is restricted to a narrower range $-0.10\,{\rm GeV} < \Delta E < 0.20\,{\rm GeV}$; this suppresses partially reconstructed $B\to D^{*}\pi$ decays that appear at lower $\Delta E$ values. The high-momentum charged pion coming directly from the $B$ meson is treated as a high-energy photon, akin to the signal mode, and is excluded from the vertex fit. We assess the systematic uncertainty due to vertex selection criteria by comparing their efficiencies for data with those for simulated events. The corresponding uncertainties are 1.0\% for channels containing a $K^{0}_{S}$ and 0.2\% for non-$K^{0}_{S}$ channels. 

We apply a similar method to evaluate the systematic uncertainties due to the $\pi^{0}/\eta$ veto and the CSBDT selection. For these studies, we use the high-statistics $B^{-}\to D^{0}[K^{-}\pi^{+}]\pi^{-}$ control sample. The high-momentum charged pion coming directly from the $B^{-}$ is treated as a high-energy photon, akin to the signal mode, and is paired with a low-energy photon from the same event to create a $\pi^{0}(\eta)$. We apply the trained $\pi^{0}/\eta$ veto to this sample and calculate the efficiency; the difference between the efficiency for data and that for simulated events is taken as the systematic uncertainty due to the veto. This value is 0.7\% for the $\pi^{0}$ veto and 0.2\% for the $\eta$ veto. The systematic uncertainty for each of the four CSBDTs trained on different $K{}^{*}\gamma$ channels is determined by applying the CSBDT to the $B^{-}\to D^{0}[K^{-}\pi^{+}]\pi^{-}$ control sample and evaluating the difference between data and simulation for the CSBDT efficiency. We assign a systematic uncertainty of 0.3--0.4\% for the application of CSBDT, depending on the channel. 

The systematic uncertainties due to possible fit bias are studied with an ensemble of datasets bootstrapped from simulated events. The systematic uncertainty is taken as the difference between the input and mean fitted values. These differences depend on the decay mode and range from 0.1\% to 0.3\%. The uncertainty due to the size of simulated samples is 0.2--0.8\%, depending on the channel. To calculate the uncertainty due to PDF modelling, the PDF shape parameters that are fixed in the fit are varied by their uncertainties. As their correlation is below 5\%, these parameters are varied independently. We refit the data many times, where for each fit the fixed parameters are sampled from Gaussian distributions and the fit result recorded. The width of the distribution of fit results is taken as the systematic uncertainty due to PDF shapes. This value is 0.1--0.4\%. The uncertainty due to modelling of the $B\overline{B}$ component is estimated by generating new PDFs using datasets bootstrapped from simulated events. This method introduces Poisson fluctuations on the PDF shape. The data are refitted with these new $B\overline{B}$ PDFs. The width of the distribution of fit results is taken as the systematic uncertainty due to $B\overline{B}$ PDF modelling. The systematic uncertainty due to the self-crossfeed fraction is assigned by varying it by $\pm50$\% and refitting the data. The uncertainty for the best candidate selection procedure is evaluated by refitting the data many times, where for each fit a signal candidate is selected randomly. The width of the distribution of fit results is taken as the systematic uncertainty. This value is 0.1--1.0\%.

The interaction of charged hadrons with detector material can lead to asymmetries in the track reconstruction efficiency.
The uncertainty due to possible mismodelling of interactions in the MC simulation is obtained from studies of $D^{0}\to K^{-}\pi^{+}$ and $D^{+}\to K^{0}_{S}\pi^{+}$ control samples.
We measure the yields of these decays for a charge-specific final state ($N$) and its $C\!P$ conjugate ($\overline{N}$) to calculate the asymmetry $\mathcal{A} = (N - \overline{N})/(N + \overline{N})$.
The asymmetry $\mathcal{A}$ can have three contributions: the $\mathcal{A}_{C\!P}$ of the control channel, a forward-backward asymmetry in the production of $D$ and $\overline{D}$ mesons due to $\gamma^{*}$--$Z^{0}$ interference in $e^{+}e^{-}\to c\overline{c}$ processes and higher-order QED effects~\cite{CCF_1, CCF_2, CCF_3}, and the instrumental asymmetry.
The forward-backward asymmetry is an antisymmetric function of the cosine of the $D$ meson polar angle in the c.m.\ frame ($\cos\theta^{*}_{D}$).
We remove this effect by averaging $\mathcal{A}$ in opposite-sign bins of $\cos\theta^{*}_{D}$. 
To obtain the instrumental asymmetry, we take $\mathcal{A}_{C\!P}$ for the Cabibbo-favored mode $D^{0}\to K^{-}\pi^{+}$ to be zero.
For $D^{+}\to K^{0}_{S}\pi^{+}$, we subtract the world average value of $\mathcal{A}_{C\!P}$~\cite{PDG} from the measured asymmetry.
The resulting values are consistent with the MC predictions within statistical uncertainties.
The effect of kinematic differences between the control samples and $B\to K^{*}\gamma$ decays is estimated by taking the difference in their $\mathcal{A}_{C\!P}$ values obtained in MC simulation.
To be conservative, we take the systematic uncertainty to be the sum in quadrature of the statistical uncertainties on the instrumental asymmetry and the $\mathcal{A}_{C\!P}$ difference.

The systematic uncertainty for the isospin asymmetry also includes a contribution from uncertainties in the $B$ lifetimes, which is around 0.2\%~\cite{PDG}. The total systematic uncertainty is obtained by summing the individual uncertainties in quadrature. The correlations between systematic uncertainties for different branching fractions are listed in Table~\ref{tab:BF_covariance}; the corresponding correlations for $\mathcal{A}_{C\!P}$ are negligible.

\begin{table}[htb!]
    \centering
    \caption{Correlations between systematic uncertainties for different branching fractions. These numbers are obtained using the lower (higher) uncertainty on the $f_{\pm}$ and $f_{00}$ values.}
    \begin{tabular}{c c c c c}
    \hline
            & $B^{0}\to K^{+}\pi^{-}\gamma$ & $B^{0}\to K^{0}_{S}\pi^{0}\gamma$ & $B^{+}\to K^{+}\pi^{0}\gamma$ & $B^{+}\to K^{0}_{S}\pi^{+}\gamma$ \\
            \hline
        $B^{0}\to K^{+}\pi^{-}\gamma$ & 1.0000 & 0.4563 (0.4194) & 0.0179 (0.1630) & 0.0359 (0.2703)\\
        $B^{0}\to K^{0}_{S}\pi^{0}\gamma$ & & 1.0000 & 0.5094 (0.6092) &  0.1629 (0.2885)\\
        $B^{+}\to K^{+}\pi^{0}\gamma$ &        &        & 1.0000 & 0.4467 (0.3678) \\
       $B^{+}\to K^{0}_{S}\pi^{+}\gamma$ &     &        &        & 1.0000 \\
    \hline
    \end{tabular}
    \label{tab:BF_covariance}
\end{table}

\section{Summary}
\label{Summary}

We report measurements of the branching fractions and $C\!P$ asymmetries for $B \to K{}^{*} \gamma$ decays. The results are
$$\mathcal{B} (B^{0} \to K{}^{*0}\gamma) = (4.14 \pm 0.10 \pm 0.11) \times 10^{-5}, $$
$$\mathcal{B} (B^{+} \to K{}^{*+}\gamma) = (4.04 \pm 0.13 {}^{+0.13}_{-0.15})\times 10^{-5},$$
$$\mathcal{B} (B \to K{}^{*}\gamma) = (4.10 \pm 0.08 \pm 0.09)\times 10^{-5},$$
$$\mathcal{A}_{C\!P} (B^{0} \to K{}^{*0}\gamma) = (-3.3 \pm 2.3 \pm 0.4)\%,$$
$$\mathcal{A}_{C\!P} (B^{+} \to K{}^{*+}\gamma) = (-0.7 \pm 2.9 \pm 0.5)\%,$$
$$\mathcal{A}_{C\!P} (B \to K{}^{*}\gamma) = (-2.4 \pm 1.9 \pm 0.3)\%.$$
In all cases, the first uncertainties are statistical and the second are systematic.
The combined results for $\mathcal{B}(B\to K^{*}\gamma)$ and $A_{C\!P}(B\to K^{*}\gamma)$ assume isospin symmetry and are obtained by taking the weighted average of the $B^{0}$ and $B^+$ results, accounting for correlations as appropriate.
We also measure the difference in $C\!P$ asymmetries between $B^{0}$ and $B^+$ and the isospin asymmetry to be
$$\Delta\mathcal{A}_{C\!P} = (+2.6 \pm 3.8 \pm 0.6)\%,$$
$$\Delta_{0+} = (+4.8 \pm 2.0 \pm 1.0 \pm 1.5)\%.$$
The last uncertainty listed for the isospin asymmetry is due to the ratio $f_{+-}/f_{00}$.
These results have comparable precision as previous measurements~\cite{BaBar_paper, LHCb_paper, Belle_paper} and are consistent with SM expectations~\cite{BSM2}. 
Our result for $\Delta_{0+}$ confirms the positive value previously reported by Belle and BABAR.

\acknowledgments

This work, based on data collected using the Belle II detector, which was built and commissioned prior to March 2019, was supported by
Higher Education and Science Committee of the Republic of Armenia Grant No.~23LCG-1C011;
Australian Research Council and Research Grants
No.~DP200101792, 
No.~DP210101900, 
No.~DP210102831, 
No.~DE220100462, 
No.~LE210100098, 
and
No.~LE230100085; 
Austrian Federal Ministry of Education, Science and Research,
Austrian Science Fund
No.~P~31361-N36
and
No.~J4625-N,
and
Horizon 2020 ERC Starting Grant No.~947006 ``InterLeptons'';
Natural Sciences and Engineering Research Council of Canada, Compute Canada and CANARIE;
National Key R\&D Program of China under Contract No.~2022YFA1601903,
National Natural Science Foundation of China and Research Grants
No.~11575017,
No.~11761141009,
No.~11705209,
No.~11975076,
No.~12135005,
No.~12150004,
No.~12161141008,
and
No.~12175041,
and Shandong Provincial Natural Science Foundation Project~ZR2022JQ02;
the Czech Science Foundation Grant No.~22-18469S;
European Research Council, Seventh Framework PIEF-GA-2013-622527,
Horizon 2020 ERC-Advanced Grants No.~267104 and No.~884719,
Horizon 2020 ERC-Consolidator Grant No.~819127,
Horizon 2020 Marie Sklodowska-Curie Grant Agreement No.~700525 ``NIOBE''
and
No.~101026516,
and
Horizon 2020 Marie Sklodowska-Curie RISE project JENNIFER2 Grant Agreement No.~822070 (European grants);
L'Institut National de Physique Nucl\'{e}aire et de Physique des Particules (IN2P3) du CNRS
and
L'Agence Nationale de la Recherche (ANR) under grant ANR-21-CE31-0009 (France);
BMBF, DFG, HGF, MPG, and AvH Foundation (Germany);
Department of Atomic Energy under Project Identification No.~RTI 4002,
Department of Science and Technology,
and
UPES SEED funding programs
No.~UPES/R\&D-SEED-INFRA/17052023/01 and
No.~UPES/R\&D-SOE/20062022/06 (India);
Israel Science Foundation Grant No.~2476/17,
U.S.-Israel Binational Science Foundation Grant No.~2016113, and
Israel Ministry of Science Grant No.~3-16543;
Istituto Nazionale di Fisica Nucleare and the Research Grants BELLE2;
Japan Society for the Promotion of Science, Grant-in-Aid for Scientific Research Grants
No.~16H03968,
No.~16H03993,
No.~16H06492,
No.~16K05323,
No.~17H01133,
No.~17H05405,
No.~18K03621,
No.~18H03710,
No.~18H05226,
No.~19H00682, 
No.~20H05850,
No.~20H05858,
No.~22H00144,
No.~22K14056,
No.~22K21347,
No.~23H05433,
No.~26220706,
and
No.~26400255,
the National Institute of Informatics, and Science Information NETwork 5 (SINET5), 
and
the Ministry of Education, Culture, Sports, Science, and Technology (MEXT) of Japan;  
National Research Foundation (NRF) of Korea Grants
No.~2016R1\-D1A1B\-02012900,
No.~2018R1\-A2B\-3003643,
No.~2018R1\-A6A1A\-06024970,
No.~2019R1\-I1A3A\-01058933,
No.~2021R1\-A6A1A\-03043957,
No.~2021R1\-F1A\-1060423,
No.~2021R1\-F1A\-1064008,
No.~2022R1\-A2C\-1003993,
and
No.~RS-2022-00197659,
Radiation Science Research Institute,
Foreign Large-Size Research Facility Application Supporting project,
the Global Science Experimental Data Hub Center of the Korea Institute of Science and Technology Information
and
KREONET/GLORIAD;
Universiti Malaya RU grant, Akademi Sains Malaysia, and Ministry of Education Malaysia;
Frontiers of Science Program Contracts
No.~FOINS-296,
No.~CB-221329,
No.~CB-236394,
No.~CB-254409,
and
No.~CB-180023, and SEP-CINVESTAV Research Grant No.~237 (Mexico);
the Polish Ministry of Science and Higher Education and the National Science Center;
the Ministry of Science and Higher Education of the Russian Federation
and
the HSE University Basic Research Program, Moscow;
University of Tabuk Research Grants
No.~S-0256-1438 and No.~S-0280-1439 (Saudi Arabia);
Slovenian Research Agency and Research Grants
No.~J1-9124
and
No.~P1-0135;
Agencia Estatal de Investigacion, Spain
Grant No.~RYC2020-029875-I
and
Generalitat Valenciana, Spain
Grant No.~CIDEGENT/2018/020;
National Science and Technology Council,
and
Ministry of Education (Taiwan);
Thailand Center of Excellence in Physics;
TUBITAK ULAKBIM (Turkey);
National Research Foundation of Ukraine, Project No.~2020.02/0257,
and
Ministry of Education and Science of Ukraine;
the U.S. National Science Foundation and Research Grants
No.~PHY-1913789 
and
No.~PHY-2111604, 
and the U.S. Department of Energy and Research Awards
No.~DE-AC06-76RLO1830, 
No.~DE-SC0007983, 
No.~DE-SC0009824, 
No.~DE-SC0009973, 
No.~DE-SC0010007, 
No.~DE-SC0010073, 
No.~DE-SC0010118, 
No.~DE-SC0010504, 
No.~DE-SC0011784, 
No.~DE-SC0012704, 
No.~DE-SC0019230, 
No.~DE-SC0021274, 
No.~DE-SC0021616, 
No.~DE-SC0022350, 
No.~DE-SC0023470; 
and
the Vietnam Academy of Science and Technology (VAST) under Grants
No.~NVCC.05.12/22-23
and
No.~DL0000.02/24-25.
S.S. acknowledges the support of SERB grant SRG/2022/001608.

These acknowledgements are not to be interpreted as an endorsement of any statement made
by any of our institutes, funding agencies, governments, or their representatives.

We thank the SuperKEKB team for delivering high-luminosity collisions;
the KEK cryogenics group for the efficient operation of the detector solenoid magnet;
the KEK computer group and the NII for on-site computing support and SINET6 network support;
and the raw-data centers at BNL, DESY, GridKa, IN2P3, INFN, and the University of Victoria for off-site computing support.


 \bibliographystyle{JHEP}
 \bibliography{biblio.bib}






\end{document}